\documentclass[aps,prl,twocolumn,showpacs,amsmath,amssymb]{revtex4}
\usepackage{epsfig} 
\usepackage{bm} 
\begin{document}
\title{Reduced and projected two-particle entanglement at finite
temperatures} \author{P. Samuelsson$^1$, I. Neder$^2$,
M. B\"uttiker$^3$} \affiliation{$^1$Division of Mathematical Physics,
Lund University, Box 118, S-221 00 Lund, Sweden}
\affiliation{$^2$Physics Department, Harvard University, Cambridge,
Massachusetts 02138, USA} \affiliation{$^3$D\'epartement de Physique
Th\'eorique, Universit\'e de Gen\`eve, CH-1211 Gen\`eve 4,
Switzerland}
 
\begin{abstract} 
We present a theory for two-particle entanglement production and detection in
mesoscopic conductors at finite temperature. The entanglement of the density
matrix projected out of the emitted many-body state differs from the
entanglement of the reduced density matrix, detectable by current correlation
measurements. Under general conditions reduced entanglement constitutes a
witness for projected entanglement. Applied to the recent experiment [Neder
{\it et al}, Nature {\bf 448} 333 (2007)] on a fermionic Hanbury Brown Twiss
two-particle interferometer we find that despite an appreciable entanglement
production in the experiment, the detectable entanglement is close to zero.
\end{abstract} 

 \maketitle

The last decade has witnessed an increasing interest in generation and
detection of entanglement in mesoscopic conductors
\cite{Beenrev,Rev1}. Entanglement is an ubiquitous quantum effect, it
describes correlations between particles that can not be accounted for
classically. A better understanding of entanglement of elementary
charge carriers, or quasiparticles, is therefore of fundamental
interest. Due to controllable system properties and coherent transport
conditions, mesoscopic conductors constitute ideal systems for the
investigation of quasiparticle entanglement. In a longer time
perspective, the prospect of quantum information processing using spin
or orbital quantum states of individual quasiparticles provides
additional motivation for such an investigation.

\begin{figure}[h]
\centerline{\psfig{figure=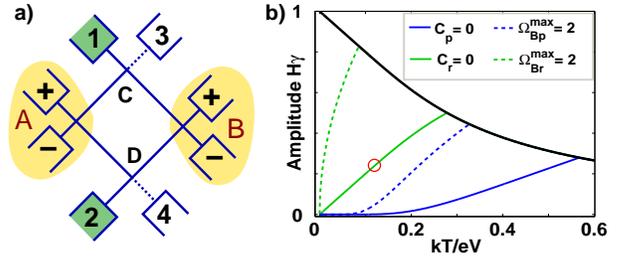,width=8.0cm}}
\caption{(color online) a) Schematic of the two-particle
interferometer (2PI) \cite{Sam04} with beam splitters C,D and biased
(grounded) contacts 1,2 (3,4). Detector regions A and B (yellow
shaded) contain beam splitters and grounded contacts $\pm$. b)
Amplitude-temperature plot for the 2PI with semi-transparent
splitters C,D. Above/to the left of the plotted lines the
entanglement of the projected (reduced) density matrix is finite,
$C_p>0~(C_r>0)$, and a Bell inequality is violated,
$\Omega_{Bp}^{max}>2 ~(\Omega_{Br}^{max}>2)$. The red ring denotes the
parameters of the experiment \cite{Neder}, showing that while
the emitted state is clearly entangled, $C_p>0$, it is barely
detectable by current and current correlation measurements, $C_r
\approx 0$.}
\label{system}
\end{figure}

To date quasiparticle entanglement has remained experimentally
elusive. However, recently an important step was taken towards a
demonstration of entanglement in mesoscopic conductors. Based on the
theoretical proposal \cite{Sam04} for a fermionic two-particle
interferometer (2PI), see Fig. \ref{system}, Neder {\it et al}
\cite{Neder} were able to demonstrate interference between two
electrons emitted from independent sources. In perfect agreement with
theory, the interference pattern was visible in the current
correlations but not in the average current. Under conditions of zero
dephasing and temperature, the part of the emitted state with one
electron in each detection region A,B would be
\begin{equation}
|\Psi_s\rangle=2^{-1/2}\left(|1\rangle_A|2\rangle_B-|2\rangle_A|1\rangle_B\right)
\label{wavefcn}
\end{equation}
where $1,2$ denote the sources. The wavefunction $|\Psi_s\rangle$ is
maximally entangled, it is a singlet in orbital, or pseudo spin,
space. However, in the experiment \cite{Neder} a reduced amplitude
($\sim 25\%$) of the current correlation oscillations was observed,
suggesting an important effect of both dephasing and finite
temperature. This raises two interesting and interrelated questions:
are the electrons reaching the detectors at A and B entangled and if
so, can this two-particle entanglement be unambiguously detected by
measurements of currents and current correlators?

In this work we provide an affirmative answer to both these questions.  We
present a general theory for two-particle entanglement generation and
detection in mesoscopic conductors at finite temperatures and apply it to the
2PI. Under very general conditions a nonzero entanglement of the reduced
orbital density matrix, accessible by current and current correlation
measurements \cite{tomo}, is shown to be a signature of finite entanglement of
the density matrix projected out from the emitted many-body state. In other
words, {\it finite reduced entanglement constitutes a witness for
finite projected entanglement}. For the 2PI-experiment, while the projected
density matrix is found to be clearly entangled, the reduced, observable
density matrix is only marginally entangled. 

We are interested in finite temperature orbital \cite{Sam03}
entanglement in a general mesoscopic system, shown in
Fig. \ref{fig3}. A theory for entanglement production in
non-interacting \cite{Been03} mesoscopic conductors at finite
temperature was presented by Beenakker \cite{Beenrev}. At a given
energy, only the component of the emitted many-body state with one
particle in detector region A and one in B has nonzero
entanglement. Formally the entanglement of the emitted state, here
called {\it projected} entanglement, is quantified in terms of the
two-particle density matrix projected out from the many-body state. In
an experiment, while this two-particle density matrix in principle can
be projected out by local operations/measurements and classical
communication between A and B, it can not be directly accessed by
standard measurements of currents and current correlators
\cite{Butrev}. The reason for this is twofold:

First, at nonzero temperatures it is not only the biased source
reservoirs which emit particles but also the grounded source
reservoirs and the detectors do. As a consequence, there is a finite
amplitude for emitted states with two-particles at A and/or at
B. These unentangled states contribute to currents and current
correlators, resulting in a detectable, effective state with
suppressed entanglement. Second, the current and current correlators
provide information on the energy integrated properties of the
many-body state but not on the emitted state at each energy. This lack
of energy-resolved information leads to a further suppression of the
detectable entanglement. Clearly, these consequences of the thermally
excited Fermi sea constitute generic problems in mesoscopic
conductors.

As a remedy for these finite temperature read-out problems it was
suggested to work with detectors at very low temperatures
\cite{Beenrev}. Recently, Hannes and Titov \cite{Titov} investigated
entanglement detection at finite temperatures via a Bell Inequality
violation. In order to overcome the problem with detectors emitting
particles they proposed to introduce energy filters, such as quantum
dots at the drains. However, these schemes \cite{Beenrev,Titov} would
lead to additional experimental complications in systems which already
are experimentally very challenging. In this work we take a different
route and investigate what information about the projected
entanglement can actually be deduced from current and current
correlation measurements. It is known \cite{tomo} that such
measurements allow for a complete, tomographic reconstruction of the
reduced orbital two-particle density matrix. At the focus of our
investigation will thus be the relation between the projected
entanglement and the entanglement of the reduced density matrix,
called the {\it reduced} entanglement.

To provide a physically compelling picture, we first investigate
entanglement generation and detection in the 2PI. Thereafter a formal
derivation for a general mesoscopic system is presented. We consider
the 2PI shown in Fig. \ref{system} with source reservoirs $1$ and $2$
biased at $eV$ while $3$ and $4$ as well as the detector reservoirs
are grounded \cite{Sam04}. All reservoirs are kept at the same
temperature $T$. The reflectionless source beam-splitters C and D have
transparencies $T_C=1-R_C$ and $T_D=1-R_D$ respectively. Note that all
electrons impinging on the detectors are emitted by the source
reservoirs; scattering between the detectors is prohibited in the
2PI-geometry.

In the 2PI-experiment, working with semi-transparent splitters $T_C=T_D=1/2$,
a two-particle Aharonov-Bohm (AB) effect \cite{Sam04} was observed in the
current cross correlations $S_{A+B+}$. For finite temperature and dephasing,
theory \cite{Vanessa} predicts
\begin{equation}
S_{A+B+}=-e^3V/(4h)H\left[1-\gamma \sin \phi_{tot}\right]
\label{noise}
\end{equation}
where $H=\coth x-1/x$ with $x=eV/2kT$, $\gamma$ is a phenomenological
decoherence parameter ranging from $1$ for a fully coherent system to
$0$ for an incoherent one and $\phi_{tot}$, up to a constant, is the
AB-phase. The applied bias $7.8\mu V$ and the estimated temperature
$10mK$ in the experiment yield $H=0.78$. A direct comparison to
Eq. (\ref{noise}) then gives the oscillation amplitude $H\gamma=0.25$,
i.e. $\gamma=0.32$, a substantial dephasing. To determine the
two-particle entanglement of the emitted state we first calculate the
projected (unnormalized) density matrix $\rho_p(E)$ at energy
$E$. Using the formal similarity of the 2PI and the reflection-less,
non spin-mixing splitter studied in \cite{Beenrev}, Eqs. (B9) - (B13),
we get
\begin{equation}
\rho_p(E)=(1-f)^2f_V^2\left[\chi
\rho_p^{diag}+(1-\chi)^2\rho^{int}\right]
\label{projected}
\end{equation}
where $\chi=e^{-2x}$ and $f=1/(1+e^{E/kT})$ and $f_V=1/(1+\chi
e^{E/kT})$ the Fermi distribution functions of the grounded and biased
source reservoirs respectively. We introduce a diagonal density matrix
$\rho_p^{diag}=\chi \hat 1\otimes \hat 1+(1-\chi)[\rho_A\otimes \hat
1+\hat 1\otimes\rho_B]$, where $\rho_A=R_C|+\rangle\langle
+|+R_D|-\rangle\langle -|$ and $\rho_B=T_C|+\rangle\langle
+|+T_D|-\rangle\langle -|$, and a density matrix
$\rho^{int}=R_CT_D|+-\rangle\langle -+|+R_DT_C|-+\rangle\langle
+-|-\gamma
\sqrt{T_CR_CT_DR_D}[e^{i\phi}|-+\rangle\langle-+|+e^{-i\phi}|+-\rangle\langle
+-|]$ resulting from the two-particle interference, with $\phi$ a
scattering phase. Here $\otimes$ is a direct product between single
particle density matrices at A and B and
$|-+\rangle\equiv|-\rangle_{Ai}|+\rangle_{Bi}$ with $\langle
+-|=(|-+\rangle)^{\dagger}$ etc. The orbital states
$|+\rangle_{Ai/Bi}$ ($|-\rangle_{Ai/Bi}$) denote the upper (lower)
incoming leads towards detector regions A/B (see Fig. \ref{fig3}). In
agreement with Eq. (\ref{noise}), decoherence $\gamma <1$, is
introduced as a suppression of the two-particle interference
$|\Psi^{int}\rangle\langle \Psi^{int}| \rightarrow \rho^{int}$, where
$|\Psi^{int}\rangle=\sqrt{R_CT_D}|+-\rangle-e^{i\phi}\sqrt{T_CR_D}|-+\rangle$.

\begin{figure}[h]
\centerline{\psfig{figure=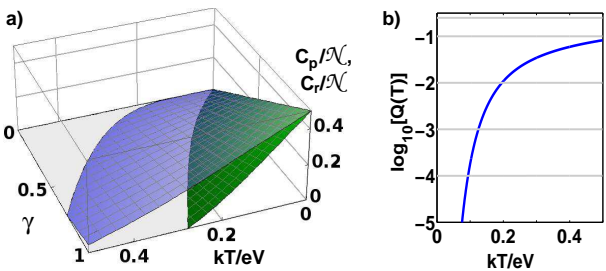,width=8.0cm}}
\caption{(color online) a) Entanglement production $C_p/{\mathcal N}$ (blue,
transparent) and $C_r/{\mathcal N}$ (green, opaque) as functions of temperature
$kT/eV$ and coherence $\gamma$ for the semi-transparent 2PI. b)
Parameter $Q$ as a function of $kT/eV$ (blue line). Values
$0.25,10^{-1},10^{-2},10^{-3},10^{-4}$ shown (grey lines). }
\label{fig2}
\end{figure}

\begin{figure}[h]
\centerline{\psfig{figure=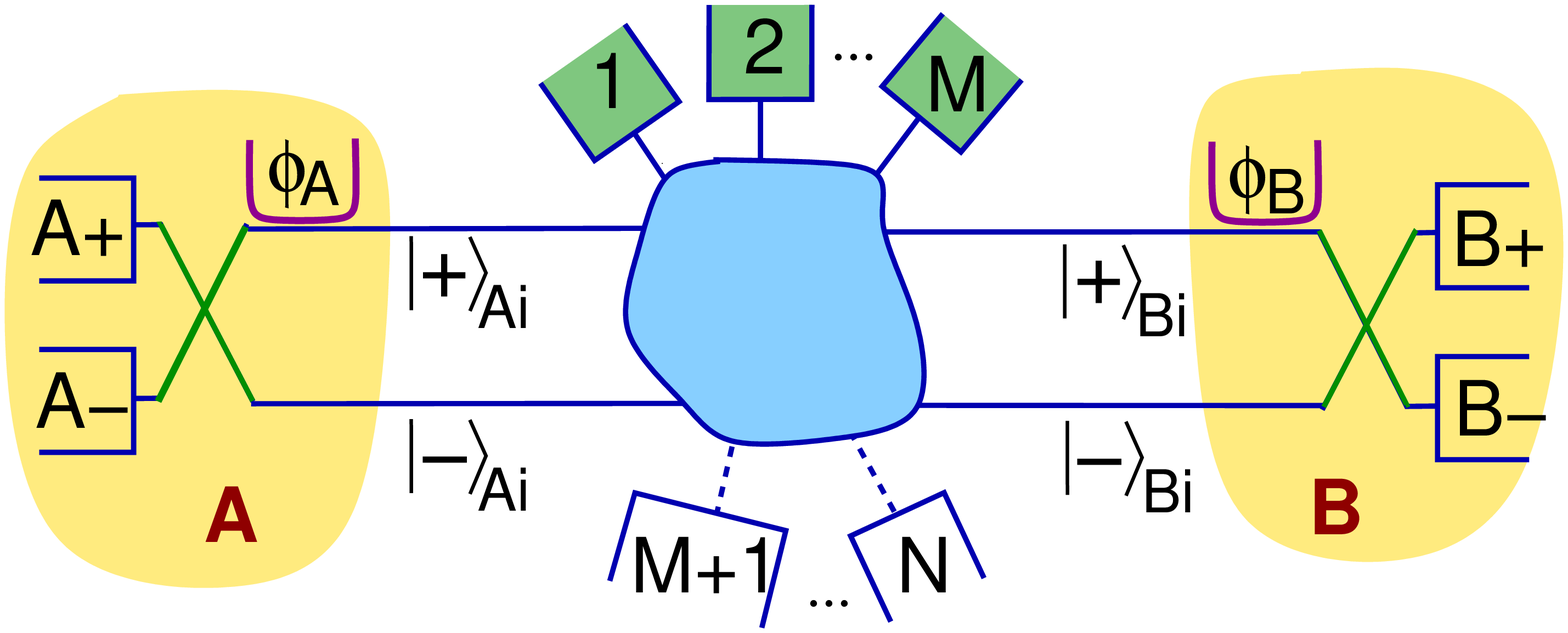,width=8.0cm}}
\caption{(color online) Sketch of the general system. A mesoscopic
scatterer is connected to $M$ biased and $N-M$ grounded source
reservoirs and, via phase gates $\phi_{A/B}$ and beam splitters, to
grounded detector reservoirs $A\pm,B\pm$. The orbital basis states
$|\pm \rangle_{Ai/Bi}$ are displayed. For details see text.}
\label{fig3}
\end{figure}

Following \cite{Beenrev} we then write $\rho_p(E)=w_p(E)\sigma_p$
where
$w_p(E)=\mbox{tr}[\rho_p(E)]=(1-f)^2f_V^2[(R_CT_D+T_CR_D)(1-\chi)^2+4\chi]$
and $\sigma_p$ the normalized, energy independent density matrix of
the emitted two-particle state. The entanglement production $C_p(E)\equiv
w_p(E)C(\sigma_p)$ is conveniently quantified in terms of the
concurrence $C$ \cite{Wooters}, ranging from $0$ for a separable state
to $1$ for a maximally entangled state. The total entanglement production
in a time $\tau$, $C_p=(\tau/h) \int dE C_p(E)$, is then
(${\mathcal N}=\tau eV/h$)
\begin{equation}
C_p=({\mathcal N}
H/2)\mbox{max}\{4\gamma\sqrt{T_CR_CT_DR_D}-\sinh^{-2}x,0\}.
\label{concproj}
\end{equation}
As shown in Fig. \ref{fig2}, $C_p$ decreases monotonically as a
function of $T$, reaching zero at a critical temperature $T_c^p$.

Inserting the parameter values from the experiment, we find that $C_p
\approx 0.1{\mathcal N}$ and $C(\sigma_p)\approx 0.3$, i.e. {\it the
state emitted by the 2PI is clearly entangled}. However, this is a
rather indirect deduction of the two-electron entanglement, an
unambiguous demonstration of entanglement would be desirable.  For
this we turn to the reduced density matrix $\rho_r$, providing full
information about the detectable two-particle correlations.

We first express $\rho_r$ in terms of currents
$I_{A\alpha},I_{B\beta}$ and low-frequency current cross correlators
\cite{Butrev} $S_{A\alpha B\beta}$, with $\alpha,\beta=\pm$. Extending
\cite{tomo} to nonzero temperatures we find
\begin{eqnarray}
&&\frac{I_{A\alpha}I_{B\beta}}{(Ve^2/h)^2}+\frac{S_{A\alpha
B\beta}}{2Ve^3/h}= \mbox{tr}\left\{\left[I_{A\alpha}^O\otimes
I_{B\beta}^O \right]\rho_r\right\}.
\label{noisecurrrel}
\end{eqnarray}
The orbital current operators in the local basis
$\{|\pm \rangle\}$, including the rotations at the detector
splitters, are $I_{A\alpha}^O=(\hat 1+\alpha S_A\sigma_zS_A^{\dagger})/2$
and $I_{B\beta}^O=(\hat 1+\beta S_B\sigma_zS_B^{\dagger})/2$, with
$\sigma_z$ the Pauli matrix and $S_A~(S_B)$ the scattering matrix of
the splitter at A (B). The results of Ref. \cite{Vanessa} yield
\begin{equation}
\rho_r=(1-H)[\rho_A\otimes\rho_B]+H\rho^{int}.
\label{reduced}
\end{equation}
Writing $\rho_r=w_r\sigma_r$ with
$w_r=\mbox{tr}[\rho_r]=[R_CT_C+R_DT_D](1-H)+R_CT_D+R_DT_C$ and
$\sigma_r$ the normalized density matrix we then define the total
entanglement production during a time $\tau$ as $C_{r}\equiv
{\mathcal N} w_rC(\sigma_r)$. It is
\begin{equation}
C_r=2{\mathcal N}\mbox{max}\{\sqrt{T_CR_CT_DR_D}[H(1+\gamma)-1],0\}.
\label{concred}
\end{equation}
As $C_p$, $C_r$ decreases monotonically with increasing $T$, reaching
zero at $T=T_c^r$, as shown in Fig. \ref{fig2}. 

Comparing Eqs. (\ref{concproj}) and (\ref{concred}) we find that
$C_p\geq C_r$ for $Q(T)=H/[4(1-H)\sinh^2 x]\leq \sqrt{T_CR_CT_DR_D}$,
independent on $\gamma$ (see Fig. \ref{fig2}). Consequently, for
splitters away from the strongly asymmetrical, tunneling limit, {\it
the reduced entanglement constitutes a lower bound for the projected
entanglement}. In the tunneling limit, however, the reduced
entanglement is larger than the projected one. This is the case since
asymmetry has different effects on $C_p$ and $C_r$; the critical
temperature $T_c^r$, from Eq. (\ref{concred}), is independent on the
splitter transparencies $T_C,T_D$ while $T_c^p$, from
Eq. (\ref{concproj}), is reduced by increasing asymmetry. For the
parameters in the experiment, $Q(T)\approx 4\times 10^{-4} \ll
\sqrt{R_CT_CR_DT_D}\approx 0.25$, showing the validity of the
bound. However, $C_r\approx 0.01{\mathcal N}$ and based on the
measurement \cite{Neder} no conclusive statement can be made about
$C_r$ and hence not about $C_p$.

A more detailed understanding of this finite temperature readout
problem can be obtained by comparing the properties of $\sigma_p$ and
$\sigma_r$. For perfect coherence $\gamma=1$ and identical beam
splitters $T_C=T_D={\mathcal T}=1-{\mathcal R}$ one can (up to a local
phase rotation) write $\sigma_{p/r}=\frac{1}{4}\xi_{p/r}\hat 1\otimes
\hat 1+(1-\xi_{p/r})|\Psi_s\rangle\langle \Psi_s|$, a Werner state
\cite{Werner}, with singlet weight \cite{Beenrev}
$1-\xi_p=2\mathcal{RT}\sinh^2x/[1+2\mathcal{RT}\sinh^2x]$ and
$1-\xi_r=H/(2-H)$. Increasing $kT/eV=2/x$ from zero, $\xi_p\approx
2e^{-2x}/(\mathcal{RT})$ becomes exponentially small while $\xi_r \approx
2/x$ increases linearly. These qualitatively different behaviors,
clearly illustrated in Fig. \ref{system}, are a striking signature of
how a small $kT/eV$, having negligible effect on $C(\sigma_p)$, leads to
a large suppression of $C(\sigma_r)$.

From Eqs. (\ref{concproj}) and (\ref{concred}) follows also a
counter-intuitive result: {\it finite amplitude of the AB-oscillations
is no guarantee for finite two-particle entanglement}. This is
apparent for $\sigma_r$ in the limit $\gamma=1$ and $T_C=T_D$, since a
separable Werner state, $\xi_r>2/3$, can be decomposed \cite{decomp}
into a sum of product states as
$\sigma_r=\frac{1}{4}\sum_{n=1}^4|\phi^A_n\rangle\langle\phi_n^A|\otimes|\phi_n^B\rangle\langle\phi_n^B|$
with $|\phi^{A/B}_n\rangle=\cos
\theta^{A/B}_n|+\rangle+e^{i\pi[1-2n]/4}\sin
\theta^{A/B}_n|-\rangle$ and
$\theta_1^{A/B}=\theta_3^{A/B}=\mbox{atan}[y^{A/B}],
\theta_2^{A/B}=\theta_4^{A/B}=-\mbox{acot}[y^{A/B}]$ and
$y^{A/B}=(\sqrt{2-\xi_r}+\sqrt{3\xi_r-2})/(\sqrt{\xi_r}\pm
\sqrt{4-3\xi_r})$ with +(-) for A(B). This {\it classically}
correlated state gives, via Eq. (\ref{noisecurrrel}), AB-oscillations
with amplitude $2(1-\xi_r)/(2-\xi_r)=H$.

Moreover, the effect of decoherence, suppressing the two-particle
interference, is similar for the projected and the reduced
entanglement. In particular, for $T=0$; $C_p=C_r$$=2{\mathcal N}\gamma
\sqrt{R_CT_CR_DT_D}$, finite for arbitrary strong dephasing
\cite{Sam04}. This is a consequence of the 2PI-geometry, prohibiting
scattering between upper (+) and lower (-) leads, i.e. pseudo-spin
flips \cite{Sam03,Turk,Sam04}. Importantly, given the controllability
of phase gates and beam splitters \cite{beamsplitters}, demonstrated
in the electronic Mach-Zehnder and 2PI experiments
\cite{phasegates,Neder}, all the technics necessary for an
entanglement test via reconstruction of $\rho_r$ are at hand.

Another widely discussed \cite{BIs,Sam03,Been03,Sam04} approach to
detect the entanglement is to use a Bell inequality. Violation of a
CHSH-Bell inequality \cite{CHSH} formulated in terms of currents and
low-frequency current correlations demonstrates finite entanglement of
$\rho_r$. An optimal CHSH-Bell test demands the same number of
measurement and level of experimental complexity as a tomographic
reconstruction of $\rho_r$. From $\sigma_p$ and $\sigma_r$, we can
using \cite{Horodecki} calculate the corresponding maximal Bell
parameters $\Omega_{Bp}^{max}$ and $\Omega_{Br}^{max}$, yielding for
identical source splitters $T_C=T_D={\mathcal T}$
\begin{eqnarray}
\Omega_{Bp/r}^{max}&=&2\sqrt{1+\gamma^2}(1-\xi_{p/r})
\end{eqnarray}
The CSHS-Bell inequalities are $\Omega_{Bp/r} \leq 2$, with the limits
$\Omega_{Bp/r}^{max}=2$ for ${\mathcal T}=1/2$ plotted in
Fig. \ref{system}. It is clear that for the values $kT/eV$ and
$\gamma$ of the 2PI-experiment, while $\Omega_{Bp} \leq 2$ in
principle can be violated, a detection of entanglement by violating
$\Omega_{Br} \leq 2$ is not possible.

Extending the above analysis to a general mesoscopic system, it can be
shown \cite{SamBut} that away from the strongly asymmetric limit (see
below), $C_r > 0$ guarantees $C_p > 0$, or equivalently $T_c^p \geq
T_c^r$, i.e. {\it for a general conductor, nonzero entanglement of
$\rho_r$ demonstrates that the emitted many-body state is
entangled}. This motivates a detailed investigation of the $\rho_r$
and $C_r$ in a general system.

We consider a conductor, shown in Fig. \ref{fig3}, characterized by a
scattering matrix $S$ and connected via single mode leads to $M\geq 2$
reservoirs biased at $eV$ and $N-M\geq 0$ grounded. The conductor is
also connected to detector reservoirs $A\alpha$ and $B\beta$ via
reflectionless splitters with controllable phase gates. The
splitter-phase gate structures perform local rotations, characterized
by $S_A$ and $S_B$, of the orbital states. All reservoirs are kept at
temperature $T$. We assume linear response in applied bias and $S$
independent on energy $E$ in the interval $-kT \lesssim E \lesssim
eV+kT$ of interest. The scattering is moreover assumed to be spin
independent and we hence drop spin notation, presenting results for a
single spin species.

In order to treat general conductors we need to modify
Eq. (\ref{noisecurrrel}) as $ S_{A\alpha B\beta} \rightarrow
S_{A\alpha B\beta}^{tr}$ where $S_{A\alpha B\beta}^{tr}\equiv
S_{A\alpha B\beta}(eV,kT)- S_{A\alpha B\beta}(0,kT)$, the
non-equilibrium excess transport part of the correlator
\cite{Butrev}. Working in the orbital basis
$\{|++\rangle,|+-\rangle,|-+\rangle,|--\rangle\}$, we get from
scattering theory \cite{Butrev} the reduced density matrix
\begin{equation}
\rho_r=t_At_A^{\dagger}\otimes t_Bt_B^{\dagger}-H
\left(t_At_B^{\dagger}\otimes t_Bt_A^{\dagger}\ \right) P,
\label{smatrho}
\end{equation}
where $P$ is a $4\times 4$ permutation matrix with nonzero elements
$P_{ij}=1$ for $ij=\{11,23,32,44\}$. The $2\times M$ transmission
matrix $t_A (t_B)$ is the matrix amplitude for scattering from the $M$
biased reservoirs to the two leads going out from the source towards
region A (B).

To obtain a compact expression for $C_r$ we make the singular value
decompositions $t_A=U_A \mathcal{T}_A V_A$ and $t_B=U_B \mathcal{T}_B
V_B$, where the $2\times M$ matrix $\mathcal{T}_A=[\tau_A,0]$ with
$\tau_A=\mbox{diag}(\sqrt{T_{A+}},\sqrt{T_{A-}})$ and similar for
$\mathcal{T}_B$, and $U_A,V_A,U_B,V_B$ unitary. Inserting this
decomposition into $\rho_r$ in Eq. (\ref{smatrho}) we arrive after
some algebra at
\begin{eqnarray}
C_r&=&2{\mathcal N}\sqrt{T_{A+}T_{A-}T_{B+}T_{B-}}~\mbox{max}\left\{F,0\right\},
\nonumber \\
F&=&H\sqrt{\zeta_1\zeta_2}-\sqrt{(1-H\zeta_1)(1-H\zeta_2)}
\label{concres}
\end{eqnarray}
where $\zeta_1,\zeta_2 \in [0,1]$ are the eigenvalues of
$ZZ^{\dagger}$, with $Z$ a $2\times2$ matrix with elements
$Z_{ij}=(V_A^{\dagger}V_B)_{ij}$ for $i,j=1,2$.

It can be shown \cite{SamBut} that for $\chi/[(1-\chi)(1-H)]<
(T_{A+}^{-1}+T_{A-}^{-1}+T_{B+}^{-1}+T_{B-}^{-1})^{-1}$, $C_r > 0$
guarantees $C_p>0$. This condition is expressed in terms
of $T_{A\pm},T_{B\pm}$, eigenvalues of the reduced single particle
density matrices accessible by a reconstruction using average currents
\cite{tomo}.

In conclusion, we have presented a theory for entanglement generation
and detection in mesoscopic conductors at finite temperatures. It is
found that under very general conditions finite reduced, detectable
entanglement constitutes a witness for nonzero emitted, projected
entanglement. The theory applied to the two-particle interferometer
\cite{Sam04} investigated experimentally by Neder et al \cite{Neder}
shows that while the emitted state is clearly entangled, the
detectable entanglement is close to zero. Our results provide
motivation for further experimental investigations of entanglement in
the 2PI.

We acknowledge discussions with E.V. Sukhorukov. The work was
supported by the Swedish VR, the Israeli SF, the MINERVA foundation,
the German Israeli Foundation (GIF) and Project Cooperation (DIP), the
US-Israel Binational SF, the Swiss NSF and MaNEP.

\end{document}